# Optical sum rule in metals with a strong interaction.

A.E. Karakozov[1], E.G. Maksimov[2]

[1] *L.F. Vereshchagin Institute for High Pressure Physics, RAS, Troitsk, Moscow region, Russia*
[2] *P.N. Lebedev Physical Institute, Moscow, Russia*

The restricted optical sum rule and its dependence on the temperature, a superconducting gap and the cutoff energy have been investigated. As known this sum rule depends on the cutoff energy and the relaxation rate $\Gamma(T)$ even for a homogeneous electron gas interacting with impurities or phonons. It is shown here that additional dependence of the spectral weight on a superconducting gap is very small in this model and this effect disappears totally when $\Gamma = 0$. The model metal with a single band is considered in details. It is well known that for this model there is the dependence of the sum rule on the temperature and the energy gap even in the case when $\Gamma = 0$. This dependence exists due to the smearing of the electron distribution function and it is expressed in the terms of Sommerfeld expansion. Here it is shown that these effects are considerably smaller than that of related with the relaxation rate if the band width is larger than the average phonon frequency. It is shown also that the experimental data about the temperature dependence of the spectral weight for the high-$T_c$ materials can be successfully explained in the framework approach based on the temperature dependence of the relaxation rate.

1. Introduction.

The general optical sum rule has been derived by Kubo[1] and can be written as

$$W = \int_0^\infty d\omega \sigma_1(\omega) = \frac{\pi}{2}\frac{ne^2}{m} \qquad (1)$$

where $\sigma_1(\omega)$ is the real part of the optical conductivity, $n$ is the total electron density and $m$ is the bare electron mass. The spectral weight $W$ does not depend on the temperature and any details of the electron structure. Recently a number of papers both experimental[2-8] and theoretical[9-15] ones have been published concerning the optical sum rule and a possible violation of this sum rule in high $T_c$ superconductors.

The real measurements of $\sigma_1(\omega)$ can be done only in a finite interval of energies up to some cutoff value $\Omega_c$. Corresponding restricted sum rule has the form

$$W(\Omega_c, T) = \int_0^{\Omega_c} d\omega \sigma_1(\omega) \qquad (2)$$

$W(T, \Omega_c)$ can be now a function of the temperature $T$, the superconducting gap $\Delta$ and the cutoff energy $\Omega_c$. Experimental investigations of high-$T_c$ superconductors[2-8] and the conventional metal gold[8] have demonstrated that the value $W(T, \Omega_c)$ depends indeed on the temperature and the cutoff energy $\Omega_c$. This dependence can be presented for the normal state in all investigated cases as[2-8]

$$W(\Omega_c, T) \approx W - B(\Omega_c)T^2 \qquad (3)$$

The coefficient $B(\Omega_c)$ decreases with the increase of the cutoff energy $\Omega_c$ but it is non equal zero even up to $\Omega_c \approx \omega_{pl}$, where $\omega_{pl}$ is the plasma frequency of electrons defined by the condition $\varepsilon_1(\omega) = 0$ and $\varepsilon_1(\omega)$ is the measured real part of the dielectric function. There is some contradictive evidence about the behavior of the spectral weight $W(\Omega_c, T)$ in the superconducting state of high temperature superconductors (HTSC) [2-8]. That concerns the conventional superconductors there is the well-known Ferrel-Glover-Tinkham (FGT) sum rule [16-18] which requires that the spectral weight lost $\Delta W$, when a metal passes from a normal to a superconducting state, must be retrieved in $\delta$ - function centered at zero frequency. This $\delta$ - function defines the contribution of the superconducting condensate. It is usually believed[18] that the FGT sum rule is satisfied in conventional superconductors at $\Omega_c \approx (4-6)\Delta$.

It is easy to understand the main peculiarities of the spectral weight $W(T, \Omega_c)$ considering the model of the homogeneous electron gas interacting with some intermediate bosons. We can write *Eq.* (2) for $W(\Omega_c, T)$ in the form

$$W(\Omega_c, T) = W - \int_{\Omega_c}^{\infty} d\omega \sigma_1(\omega) \qquad (4)$$

It follows from this expression that the properties of the restricted sum rule depend on the behavior of the conductivity at considerably high energies. It is easy to show[13] that for $\omega \gg (2\Delta, \overline{\omega})$ ($\overline{\omega}$ is some average boson frequency) the conductivity $\sigma_1(\omega)$ can be written as

$$\sigma_1(\omega) = \sigma_1^N(\omega)\left(1 - \alpha\frac{\overline{\Delta}^2}{\omega^2}\right) \qquad (5)$$

This expression can be derived in the framework of the Eliashberg equation[19] for isotropic pairing as well as for anisotropic pairing due to spin fluctuations[20]. Here the numerical coefficient $\alpha$ is the order of the unity and it is included to take into account the possible averaging of an angular dependence of the gap function and $\overline{\Delta}^2$ is some average square gap. The conductivity of the normal state $\sigma_1^N(\omega)$ can be written in the considered model as[13]

$$\sigma_1^N(\omega) = \frac{\omega_{pl}^2}{4\pi}\frac{2\Gamma(\omega = \infty, T)}{\omega^2 + (2\Gamma(\omega = \infty, T))^2} \qquad (6)$$

Here $\Gamma(\omega = \infty, T)$ is the relaxation rate due to the interaction of electrons with bosons. Using *Eqs.*(4-6) we can write expression for the optical weight as

$$W(\Omega_c, T) = \frac{\omega_{pl}^2}{8}\left[1 - \frac{2}{\pi}\left(\frac{2\Gamma(\omega = \infty, T)}{\Omega_c}\right)\left(1 - \frac{2}{3\pi}\alpha\frac{\overline{\Delta}^2}{\Omega_c^2}\right)\right] \qquad (7)$$

This expression shows that the FGT sum rule is satisfied in the absence of the relaxation in the system such as BCS model at any value of $\Omega_c$. Moreover, we see that there is no any direct contribution to the FGT sum rule from the superconducting gap. This contribution is multiplied by the factor $\frac{\Gamma}{\Omega_c}$ which has usually additional smallness. It was shown in our preceding work[13] that the relaxation rate for discussed system demonstrates the quadratic dependence on the temperature in the interval temperatures $100K \leq T \leq 300K$. This temperature dependence of the deviation of the spectral weight from the value $W$ well described the experimental data obtained for convention metal $Au$ and even for high-$T_c$ superconductors. The main part of the theoretical works concerning the restricted optical sum rule has been based on the consideration not of the homogeneous electron gas but for one band model. The main goal of this our work is to study in the details the restricted optical sum rule for a one band model where there is a strong interaction of electrons with some low energy bosons.

2. The deviation of basic equations.

As we have mentioned before the restricted optical sum rule has been investigated recently for band electrons. Moreover, it was proposed that there is one conducting band which is well separated from other bands by a rather large interband gap $E_g$ which is larger than the corresponding band width ($2W_b$). The Hamiltonian of such system in the presence of uniform electromagnetic field can be written as

$$H = \sum_{\mathbf{p},\sigma} \xi\left(\mathbf{p} - \frac{e}{c}\mathbf{A}\right) a^+_{\mathbf{p}\sigma} a_{\mathbf{p}\sigma} + \sum_{\mathbf{p},\sigma,\mathbf{q},\lambda} g(\mathbf{q},\lambda) a^+_{\mathbf{p}+\mathbf{q}\sigma} a_{\mathbf{p}\sigma} \left(b_{\mathbf{q}\lambda} + b^+_{-\mathbf{q}\lambda}\right) + H_{inter} \tag{8}$$

Here $\mathbf{A}$ is the electromagnetic potential and $a^+_{\mathbf{p}\sigma}$, $a_{\mathbf{p}\sigma}$ are creation and annihilation operators for electrons, $g(\mathbf{q},\lambda)$ is the matrix element of electron-boson interaction and $b^+_{\mathbf{q}\lambda}$, $b_{\mathbf{q}\lambda}$ are bosons creation and annihilation operators. The operator $H_{inter}$ presents interband transitions. The role of interband transitions in the system with a strong electron-phonon interaction (EPI) has been discussed by Holstein[21]. It was shown that the interband transitions can be neglected in the first approximation if the interband gap $E_g$ is larger than the band width $W_b$ and all characteristic energies of the electron-phonon system.

We can expand the kinetic term of the Hamiltonian (8) as a power series in $\mathbf{A}$ and get

$$H_{kin} = \sum_{\mathbf{p},\sigma} \xi(\mathbf{p}) a^+_{\mathbf{p}\sigma} a_{\mathbf{p}\sigma} - \frac{e}{c} \sum_{\mathbf{p},\sigma} \frac{\partial \xi(\mathbf{p})}{\partial p_\alpha} a^+_{\mathbf{p}\sigma} a_{\mathbf{p}\sigma} A_\alpha + \frac{e^2}{2c^2} \sum_{\mathbf{p},\sigma} \frac{\partial^2 \xi(\mathbf{p})}{\partial p_\alpha \partial p_\beta} a^+_{\mathbf{p}\sigma} a_{\mathbf{p}\sigma} A_\alpha A_\beta \tag{9}$$

Now we can write the current operator as

$$j_\alpha = -c \frac{\delta H_{kin}}{\delta A_\alpha} = e \sum_{\mathbf{p},\sigma} \frac{\partial \xi(\mathbf{p})}{\partial p_\alpha} a^+_{\mathbf{p}\sigma} a_{\mathbf{p}\sigma} - \frac{e^2}{c} \sum_{\mathbf{p},\sigma} \frac{\partial^2 \xi(\mathbf{p})}{\partial p_\alpha \partial p_\beta} a^+_{\mathbf{p}\sigma} a_{\mathbf{p}\sigma} A_\beta \tag{10}$$

The first term on the right side of *Eq.* (12) is the paramagnetic current and the second one is the diamagnetic contribution. By evaluating $\langle j_\alpha \rangle$ in linear response[21,22] and taking its Fourier components, we obtain the complex optical conductivity

$$\sigma_{\alpha\alpha}(\omega) = \frac{ie^2}{\omega + i\delta} \left[K^d_{\alpha\alpha} + \Pi_{\alpha\alpha}(\omega)\right] \tag{11}$$

Here $K^d$ diamagnetic response kernel and $\Pi_{\alpha\alpha}(\omega)$ is the current-current correlation function. The real part of the optical conductivity can be written as

$$\sigma_1(\omega) = e^2 \pi \left[K^d + \Pi(0)\right] \delta(\omega) - e^2 \frac{\mathrm{Im}\,\Pi(\omega)}{\omega} \tag{12}$$

The *Eqs.*(11) and (12) are written in the assumption that the system under consideration has a cubic symmetry but it is not so important for the future discussion and we shall omit coordinate indexes below. The optical sum rule can be written now as

$$\int_0^{\Omega_c} d\omega \, \sigma_1(\omega) = e^2 \frac{\pi}{2} \left[K^d + \Pi(0)\right] - e^2 \int_0^{\Omega_c} d\omega \frac{\mathrm{Im}\,\Pi(\omega)}{\omega} \tag{13}$$

If we consider the case when the cutoff energy $\Omega_c$ is larger than the band width $2W_b$ but smaller than $E_g$ we can change the upper limit in *Eq.* (13) on $\Omega_c \to \infty$. Remembering, that the current-current correlation function $\Pi(\omega)$ satisfies the Kramers-Kronig relation

$$\Pi(\omega) = \frac{2}{\pi} \int_0^\infty d\omega' \omega' \frac{\operatorname{Im}\Pi(\omega')}{\omega'^2 - \omega^2 - i\omega\delta} \tag{14}$$

we shall immediately get that

$$W(T) = \int_0^{\Omega_c} d\omega \sigma_1^\alpha(\omega) = e^2 \frac{\pi}{2} K^d = e^2 \pi \sum_\mathbf{p} \frac{\partial^2 \xi_\mathbf{p}}{\partial p_\alpha^2} n_\mathbf{p} \tag{15}$$

This answer is true for the case $\Omega_c > W_b$ both for normal and superconducting states. Here $n_\mathbf{p}$ is the electron distribution function which should be calculated taking into account the interaction part of the Hamiltonian (8) related with the electron-phonon interaction (EPI).

The mean part of works[9-12,14,15] about the restricted optical sum rule used namely *Eq.* (15) for all discussions about the violation of the optical sum rules and their dependence on the temperature, superconducting gap, etc.. We should mentioned , however, that the experimental data demonstrated unambiguously the dependence of the spectral weight on the cutoff energy up to considerably high values of $\Omega_c$ both in the ordinary metals[8] and in high-$T_c$ systems[2-8]. We shall return to the discussion of this problem a little later but now we consider the case when $\Omega_c > W_b$.

Firstly, we would like to mention that for a normal state of metals there is the identity

$$K^d \equiv -\Pi(0) \tag{16}$$

It can be proved in the general case using the Green functions approach and the Word identity[15,22]. We shall not do that here but we shall show below that the identity (16) is satisfied in our one band model with isotropic EPI. Taking in to account *Eq.* (16) $\sigma_1(\omega)$ can be written as

$$\sigma_1(\omega) = -e^2 \frac{\operatorname{Im}\Pi(\omega)}{\omega} \tag{17}$$

In the framework of the thermodynamical theory of perturbations the function $\Pi(i\omega_n)$ has the form

$$\Pi(i\omega_n) = 2 \sum_\mathbf{p} \left(\frac{\partial \xi_\mathbf{p}}{\partial \mathbf{p}}\right)^2 T \sum_{\omega_m} G(\mathbf{p}, i\omega_m + i\omega_n) G(\mathbf{p}, i\omega_m) \tag{18}$$

where $T$ is temperature and $\omega_n = \pi T(2n+1)$. The factor 2 in this expression is the result of the summation over the spins. Here $G(\mathbf{p}, i\omega_n)$ is the electron Green function which is equal

$$G^{-1}(\mathbf{p}, i\omega_n) = i\omega_n - \xi_\mathbf{p} - \Sigma(i\omega_n) \tag{19}$$

where $\Sigma(i\omega_n)$ is the electron self energy which should be calculated using the Hamiltonian (8). It is well known that the vertex correction function can be neglected for isotropic EPI. The expression (18) for the $\Pi(i\omega_n)$ can be easily analytically continued on the real $\omega$ axis using the electron spectral density

$$A(\mathbf{p}, \omega) = -\frac{1}{\pi} \operatorname{Im} G(\mathbf{p}, i\omega_n \to \omega + i\delta) \tag{20}$$

and the electron Green functions spectral representations[22]. Then one can obtain

$$\sigma_1(\omega) = e^2 \frac{2\pi}{\omega} \sum_\mathbf{p} \left(\frac{\partial \xi_\mathbf{p}}{\partial \mathbf{p}}\right)^2 \int_{-\infty}^\infty d\omega' [f(\omega') - f(\omega' + \omega)] A(\mathbf{p}, \omega') A(\mathbf{p}, \omega' + \omega) \tag{21}$$

Here $f(\omega)$ is the Fermi distribution function. The electron distribution function $n_\mathbf{p}$ in *Eq.* (15) can be also expressed in terms of $A(\mathbf{p},\omega)$

$$n_\mathbf{p} = \int_{-\infty}^{\infty} d\omega f(\omega) A(\xi_\mathbf{p},\omega) \qquad (22)$$

Now we show that the identity (16) is satisfied exactly in our model. We shall write for this goal $\Pi(0)$ as

$$\Pi(0) = \frac{2}{\pi} \int_{0+}^{\infty} d\omega \frac{\operatorname{Im}\Pi(\omega)}{\omega} =$$

$$= -4 \sum_\mathbf{p} \left(\frac{\partial \xi_\mathbf{p}}{\partial \mathbf{p}}\right)^2 \int_{0+}^{\infty} \frac{d\omega}{\omega} \int_{-\infty}^{\infty} d\omega' [f(\omega') - f(\omega'+\omega)] A(\mathbf{p},\omega') A(\mathbf{p},\omega'+\omega) \qquad (23)$$

Using the electron Green functions spectral representations this expression can be reduced to

$$\Pi(0) = 2 \sum_\mathbf{p} \left(\frac{\partial \xi_\mathbf{p}}{\partial \mathbf{p}}\right)^2 \int_{-\infty}^{\infty} d\omega' f(\omega') \operatorname{Im} G^2(\xi_\mathbf{p},\omega') \qquad (24)$$

and taking into account the *Eq.* (19), one obtains

$$\Pi(0) = 2 \sum_\mathbf{p} \left(\frac{\partial \xi_\mathbf{p}}{\partial \mathbf{p}}\right)^2 \int_{-\infty}^{\infty} d\omega' f(\omega') \frac{\partial}{\partial \xi_\mathbf{p}} A(\xi_\mathbf{p},\omega') \qquad (25)$$

The next step consists in integrating per part on the momentum $\mathbf{p}$ and get

$$\Pi(0) = -2 \sum_\mathbf{p} \frac{\partial^2 \xi_\mathbf{p}}{\partial \mathbf{p}^2} n_\mathbf{p} \qquad (26)$$

Comparing *Eq.* (26) and *Eq.* (15) we see that the identity (16) is indeed satisfied exactly in the considered model.

Before proceeding to the detailed investigations of the restricted optical sum rule in one band approximation we should mention one distinction of this model from the homogeneous electron gas. As it was pointed by van der Marel[10,11] the spectral weight (15) can depend on the temperature both in the normal and superconducting states even in absence of any relaxation. We can write the expression for $W(T)$ using *Eq.* (15) as

$$W(T) = \pi e^2 \sum_\mathbf{p} \frac{\partial^2 \xi_\mathbf{p}}{\partial \mathbf{p}^2} f_\mathbf{p} \qquad (27)$$

Here $f_\mathbf{p}$ is the Fermi distribution function for normal state

$$f_\mathbf{p} = \left(e^{\frac{\xi_\mathbf{p} - \mu}{T}} + 1\right)^{-1} \qquad (28)$$

($\mu$ is the chemical potential). Correspondingly, for a superconducting state the distribution function has the form[22]

$$f_\mathbf{p} = -\frac{1}{\pi} \int_{-\infty}^{\infty} d\omega f(\omega) \operatorname{Im} G^{11}_{BCS}(\omega) = \frac{1}{2}\left[1 - \frac{\xi_\mathbf{p} - \mu}{E_\mathbf{p}} \tanh \frac{E_\mathbf{p}}{2T}\right] \qquad (29)$$

where $G^{11}_{BCS}(\omega)$ is the diagonal matrix element of the electron Green function in BCS approximation and $E_\mathbf{p} = \sqrt{(\xi_\mathbf{p} - \mu)^2 + \Delta^2}$ is the quasiparticle dispersion in the superconducting state. The temperature dependence of $W(T)$ in the normal state comes from the temperature smearing of the Fermi function and it can be easily evaluated using so-called Sommerfeld expansion. The simple estimation gives us

$$W(T) = W(0)\left(1 - \beta \frac{T^2}{W_b^2}\right) \tag{30}$$

where $\beta$ is some numerical coefficient. This result resembles with the experimental observation of the temperature dependence of $W(T)$. However, as it was emphasized in the works[8,15] that there is only the quantitative analogy between the experimental data and the result obtained in one band approximation for noninteracting electrons. The measured value of the coefficient $B$ characterizing the temperature dependence of $W(T)$ is at least one order of magnitude larger than expected from *Eq.* (30). This result have been obtained in the works[8,15] using the value of the band width $2W_b$ taking from the ARPES measurements of the Fermi surface in cuprates. This approach, from our point of view[23], leads to underestimating the value $W_b$. It means that the real disagreement between experimental data about $W(T)$ and noninteracting electron result for one band model could be even larger. The temperature dependence of $W(T)$ in a superconducting state is related with the smearing of the distribution function (*Eq.* (29)) with the increasing of the superconducting gap $\Delta$. It results to decrease $W(T)$ with decreasing $T$ in superconducting state. The absolute value of this decreasing of $W(T)$ is also small as the ratio $\left(\frac{\Delta}{W_b}\right)^2$. We shall show below that the effects existing due to the change of the relaxation rate can be much large than which have been discussed above even at $T < T_c$.

### 3. Electron-phonon interaction in one-band model.

Using the Hamiltonian (8) and neglecting interband transitions the equation for the electron self-energy $\Sigma(i\omega_n)$ can be writing as[24]

$$\Sigma(i\omega_n) = \pi T \sum_{m,\mathbf{k}} \left[\lambda(i\omega_n - i\omega_m) + \delta_{nm}\frac{\gamma_{imp}}{\pi T}\right]\left\{-\frac{1}{\pi}\int_{-\infty}^{\infty}\frac{d\xi}{N_b(\mu)}\delta(\xi - \xi_\mathbf{k})G(\xi_\mathbf{k}, i\omega_n)\right\} \tag{31}$$

$$\lambda(i\omega_n) = \int_{-\infty}^{\infty} d\Omega \frac{\Omega \alpha^2(\Omega)F(\Omega)}{\Omega^2 - (i\omega_n)^2} \tag{32}$$

$$N_b(\xi) = \sum_\mathbf{k} \delta(\xi - \xi_\mathbf{k}) \tag{33}$$

where $\gamma_{imp}$ is the relaxation rate due to electron scattering on impurities, $\alpha^2(\omega)F(\omega)$ is the Eliashberg spectral function and $N_b(\xi)$ is the bare band electron density of states. In a superconducting state $\Sigma(i\omega_n)$ and Green's function $G(\xi_\mathbf{k}, i\omega_n)$ are matrices in the Nambu matrixes space. Further we shall present explicitly the expressions for normal state only. The analytical continuation *Eq.* (31) on the real axis for this case can be written as

$$\omega[Z(\omega) - 1] = \frac{\pi}{2}\int_{-\infty}^{\infty} dz\left\{th\frac{z}{2T}\Lambda(\omega - z)\text{Im}\,g(z) + cth\frac{\omega - z}{2T}\text{Im}\,\Lambda(\omega - z)g(z)\right\} + \frac{\gamma_{imp}}{N_b(\mu)}g(\omega) \tag{34}$$

where $\omega[Z(\omega) - 1] = -\Sigma(\omega)$ and

$$g(\omega) = -\frac{1}{\pi}\sum_\mathbf{p}\frac{1}{N_b(\mu)}G(\xi_\mathbf{p}, \omega + i\delta) \tag{35}$$

$$\Lambda(\omega) = -\frac{1}{\pi}\int_{-\infty}^{\infty} d\Omega \frac{\alpha^2(\Omega)F(\Omega)}{\omega+i\delta-\Omega} \equiv K_{ph}(\omega) + i\alpha^2(\omega)F(\omega) \tag{36}$$

The imaginary part of the function $g(\omega)$ gives the electron density of states in the interacting system. Usually the function $\omega Z(\omega)$ is presented in the form

$$\omega Z(\omega) = \omega[1+\lambda(\omega)] + i\Gamma(\omega) \tag{37}$$

where $\lambda(\omega)$ and $\Gamma(\omega) = \tau^{-1}(\omega)$ define the mass renormalization and the relaxation rate, correspondingly. The constant of EPI $\lambda$ is expressed as $\lambda = \pi K_{ph}(0)$. *Eqs.* (34) - (35) should be solved selfconsistently. This procedure is closely similar with that of used in coherent potential approximation (CPA) in the theory of disordered metals[25]. For not too narrow bands $\left(\Omega_{ph}/\overline{\xi} \ll 1\right)$ *Eq.* (34) can be easily solved at energies $\omega \ll \overline{\xi}$ where $\Omega_{ph}$ and $\overline{\xi}$ are characteristic energies of phonons and electrons, correspondingly, and ($\overline{\xi} \sim \varepsilon_F, W_b$). In this energy region $N_b(\xi) \approx N_b(\mu)$ and the function $g(\omega)$ can be changed with accuracy $\left|\omega Z(\omega)/\overline{\xi}\right|$ on the bare band function $g_b(\omega)$ or in the limit $W_b \to \infty$ ($\varepsilon_F/W_b \to 0$) on the free electron gas function $g_0(\omega) \approx i$. Now the solution can be obtained in the form which is well known from the investigations of the model of homogeneous electron gas[21]

$$\omega[Z_0(\omega)-1] = \frac{\pi}{2}\int_{-\infty}^{\infty} dz \left\{ th\frac{z}{2T}K_{ph}(\omega-z) + i\left[cth\frac{\omega-z}{2T} + th\frac{z}{2T}\right]\alpha^2 F(\omega-z) \right\} \tag{38}$$

The behavior of the function $Z_0(\omega)$ is also well known and can be presented at $\omega, T \ll \Omega_{ph}$ as

$$\omega Z_0(\omega) \approx \omega(1+\lambda) + i\lambda \frac{T^3}{\Omega_{ph}^2} \to \omega(1+\lambda) + i\delta \tag{39}$$

It leads to renormalization of the bare band effective mass $m_b$

$$m_0^* = m_b(1+\lambda) \tag{40}$$

and, correspondently, to increasing of the low temperature electron specific heat

$$c_e = \frac{\pi^2}{3} T N_b(\mu)(1+\lambda) \tag{41}$$

The main feature of the one-band model with EPI can be understood from the solutions of *Eq.* (34) for the cases of the impurity scattering and for EPI at high temperatures

$$\left(\frac{\Omega_{ph}}{2\pi T}\right)^2 \ll 1 \tag{42}$$

To find these solutions we shall use the Hubbard ellipse model[25] for the band electron density of states per spin

$$N_b(\xi) = \frac{2}{\pi W_b}\sqrt{1 - \left(\frac{\xi - W_b}{W_b}\right)^2} \tag{43}$$

which has been used successfully in the investigations of the disordered metals[26].

In accordance with condition (42) $\lambda(i\omega_n - i\omega_m) \sim \delta_{nm}$ *Eq.* (34) has the form

$$\Sigma(\omega) = \Gamma\frac{1}{\pi}\int_{-\infty}^{\infty} d\xi \frac{N_b(\xi)}{N_b(\mu)} G(\xi,\omega) \equiv -\frac{\Gamma}{N_b(\mu)} g(\omega) \tag{44}$$

The electron interaction is characterized in *Eq.* (44) by a single parameter $\Gamma$ which is equal $\gamma_{imp}$ for the elastic scattering on impurities and is equal $\lambda \pi T$ for the EPI. We consider metals with the electron-hole symmetry which means that $\varepsilon_F = \mu(T=0) = W_b$ and the chemical potential $\mu$ does not depend on the temperature. The band function $g_b(z)$ for any complex argument $z$ can be written as

$$g_b(z) = -\frac{1}{\pi}\int_{-\infty}^{\infty} d\xi \frac{N_b(\xi)}{z-\xi} = -\frac{1}{W_b} N_b(\mu = W_b)\left(z - \sqrt{z^2 - W_b^2}\right) \quad (45)$$

The expression for the function $g(\omega)$ has now the simple form $g(\omega) = g_b(\omega - \Sigma(\omega))$ which leads to the expression for $\Sigma(\omega)$ as

$$\Sigma(\omega) = \frac{\Gamma}{W_b}\left(\omega - \Sigma(\omega) - \sqrt{(\omega - \Sigma(\omega))^2 - W_b^2}\right) \quad (46)$$

Solving this equation we get

$$\Sigma(\omega) = \frac{\beta}{1+2\beta}\left(\omega - \sqrt{\omega^2 - W_b^2(1+2\beta)}\right) \quad (47)$$

$$g(\omega) = -\frac{1}{W_b} N_b(W_b) \frac{1}{1+2\beta}\left(\omega - \sqrt{\omega^2 - W_b^2(1+2\beta)}\right) \quad (48)$$

where $\beta = \Gamma/W_b$. We can also write the expression for the renormalization function $Z(\omega)$ as

$$\omega - \Sigma(\omega) = \omega Z(\omega) = \frac{1}{1+2\beta}\left[(1+\beta)\omega + \beta\sqrt{\omega^2 - W_b^2(1+2\beta)}\right] \quad (49)$$

It is easy to see from the *Eq.* (49) that the electron mass determined by elastic scattering or EPI $m_{el}^*$ decreases as

$$m_{el}^* \approx m_b\left(1 - \frac{\Gamma}{W}\right) \quad (50)$$

This result comes from the fact that the elastic scattering of electrons in a finite band leads to the expansion of the band due to a finite electron life time $\tau(\xi_\mathbf{p})$. It leads in its turn to the smearing of the electron density of states which diminishes the electron mass as

$$m_{el}^*(T) \sim \frac{W_b}{W_b + \Gamma(T)} m_b \approx m_b\left(1 - \frac{\Gamma(T)}{W_b}\right) \quad (51)$$

where $\Gamma(T)$ is of order of the maximum of the relaxation rate for the band electrons. This behavior is very distinct from that of obtained for the mass renormalization determined by inelastic EPI at low energies and temperatures ( *Eq.* (40)).

The result given by *Eq.* (50) was obtained from the approximation solutions *Eqs.* (34), (35) at the condition (42). We show now that the approximate solution for the selfenergy (44) can be applied in the case $\left(\Omega_{ph}/W_b\right) \ll 1$ even for all temperatures using the special expression for the relaxation rate

$$\Gamma(T) \approx \Gamma(\infty, T) = \pi\int_0^{\infty} d\omega \alpha^2 F(\omega) cth\left(\frac{\omega}{2T}\right) \quad . \quad (52)$$

We have confirmed this fact by the numerical solutions *Eqs.* (34), (35) for different forms of Eliashberg functions.

On *Figs.* 1-3 the results of this selfconsistent solution are shown. We have used for the calculations shown on *Figs.* 1, 2 the function $\alpha^2(\omega)F(\omega)$ taking from the work[13] with $\lambda \approx 1.5$., the average phonon frequency $\Omega_{ph} \approx 300K$, the band width $2W_b = 2*10^4 sm^{-1}$, and $\gamma_{imp} = 0$.

The approximate solution and the result given by *Eq.* (40) are also shown on *Figs.* 1, 2. It can be seen that the numerical solution coincides considerably well with that of given by *Eq.* (40) at low energies $\omega < \Omega_{ph}$ and with the approximate solution (*Eqs.* (44) - (48), (52)) for large energies. The *Fig.* 3 demonstrates the mass renormalization at low temperatures on whole energy interval. It is clearly seen from this *Fig.* that there is the change of the sign of the mass renormalization at $\omega \approx \Omega_{ph}$.

We do not reproduce here the results of calculations for the superconducting state. They are not different from that of obtained for the homogeneous model because the superconducting energy gap $\Delta(\omega)$ is important only at considerably low energies $(\Delta(\omega) << \bar{\xi})$ and coincides with the solution for normal state for larger energies.

### 4. Optical spectral weight in one-band model with EPI.

The spectral weight defined by the Eq.(15) can be rewritten in the form

$$W(T) = \pi e^2 \int_{-\infty}^{\infty} d\omega f(\omega) \left\{ -\frac{1}{\pi} \mathrm{Im} \int_{-\infty}^{\infty} d\xi G(\xi, \omega) \right\} \frac{\partial}{\partial \xi} \sum_{\mathbf{p}} v_{\mathbf{p}}^x v_{\mathbf{p}}^x \delta(\xi_{\mathbf{p}} - \xi) \quad (53)$$

We shall use for the presentation of the value $\sum_{\mathbf{p}} v_{\mathbf{p}}^x v_{\mathbf{p}}^x \delta(\xi_{\mathbf{p}} - \xi)$ the model function

$$\sum_{\mathbf{p}} v_{\mathbf{p}}^x v_{\mathbf{p}}^x \delta(\xi_{\mathbf{p}} - \xi) = \langle V_x^2(\xi) \rangle N_b(\xi) = V_x^2 \frac{2}{\pi W_b} \left[ 1 - \left( \frac{\xi - W_b}{W_b} \right)^2 \right]^{\frac{3}{2}} \quad (54)$$

which is directly connected with the representation of the density of states (*Eq.* (43)). The value of $V_x^2$ in *Eq.* (54) is determined as $V_x^2 = \frac{1}{3} V_m^2$ where $V_m$ is the maximal value of electron velocity in the band. Using the approximation (54) we obtain the spectral weight in the form

$$W(T) = 3 \frac{\tilde{\omega}_b^2}{W_b^2} \int_0^\infty d\omega \mathrm{Im} \left\{ \frac{\omega Z(\omega)}{N_b(W_b)} g(\omega) \right\} - 6 \frac{\tilde{\omega}_b^2}{W_b^2} \int_0^\infty d\omega f(\omega) \mathrm{Im} \left\{ \frac{\omega Z(\omega)}{N_b(W_b)} g(\omega) \right\} \quad (55)$$

where $\tilde{\omega}_b^2 = \pi e^2 V_x^2 N_b(W_b) = \frac{\omega_{pl}^2}{8}$ and $\omega_{pl}$ is the electron plasma frequency for a given band.

The first term of the *Eq.* (55) depends on the temperature only due to interaction electrons with phonons. The second term gives the Sommerfeld contribution to the spectral weight. Using the approximation of the elastic scattering (*Eqs.* (47, 48, 52)), we can calculate the spectral weight exactly and obtain

$$W(T) = \frac{\tilde{\omega}_b^2}{\sqrt{1 + 2\beta}} - \frac{\tilde{\omega}_b^2}{(1 + 2\beta)^{\frac{3}{2}}} \frac{\pi^2}{2} \frac{T^2}{W_b^2} \quad (56)$$

It can be presented for a small value of $\beta$ as

$$\frac{W(T)}{\tilde{\omega}_b^2} \approx 1 - \frac{\Gamma}{W_b} - \frac{\pi^2}{2} \frac{T^2}{W_b^2} \quad (57)$$

Using for $\Gamma(T)$ even simplest estimation given by *Eq.* (39) one can see that the contribution related with relaxation rate (the second term in the right hand of *Eq.* (57)) becomes to be larger than the Sommerfeld contribution (the third term) when

$$\frac{T}{\Omega_{ph}} \geq \alpha(\lambda) \frac{\Omega_{ph}}{W_b} \quad (58)$$

here $\alpha(\lambda)$ for the case of a strong EPI is the numerical coefficient of the order of the unity. Let us note also, that the spectral weight even at $T = 0$ is not equal $\frac{\omega_b^2}{8}$ and approaches to this value in the limit $W_b \to \infty$ only.

Now we shall discus the temperature dependence of the spectral weight described by *Eqs.* (56), (57). The temperature dependence of $\Gamma(\infty, T)$ depends on the details of the form of the Eliashberg spectral function $\alpha^2(\omega)F(\omega)$. It is well known that temperature dependent part of $\Gamma(\infty, T)$ can be presented as $\sim T^\alpha$ where $\alpha \geq 3$ at low temperature $T \leq \Omega_{ph}$ and $\sim T$ at $T \gg \Omega_{ph}$. It was shown in our preceding work[13] that the temperature dependence of $\Gamma(\infty, T)$ at the interval of intermediate temperatures is very close to $T^2$ in a homogeneous electron gas. The rough notion about the temperature dependence of $\Gamma(\infty, T)$ can be obtained from *Figs.* 4*a* and 4*b*. These figures show the function

$$w_{\text{mod}}(T) = 1 - \frac{\Gamma(T)}{W_b} \approx 1 - \frac{\pi}{2} \frac{\lambda \Omega_E}{W_b} - \pi \frac{\lambda \Omega_E}{W_b} n_B\left(\frac{\Omega_E}{T}\right) \tag{59}$$

where $n_B$ is the Bose function and $\Gamma(T)$ is the relaxation rate for the Einstein phonon spectra with phonon frequencies $\Omega_E = 200K$ (*Fig.* 4*a*) and $\Omega_E = 400K$ (*Fig.* 4*b*). We have used the EPI constant of coupling $\lambda = 1.5$ and $W_b \approx 2ev$ for both cases. The curves demonstrate $T^2$ behavior on the temperature interval $100K \leq T \leq 200K$ but they deviate from $T^2$ behavior at low temperatures as it can be seen on *Figs.* 4*a*, 4*b* and at high $T$ (which is not shown on *Figs.*). We would like to mentioned here that *Fig.* 4*a* reproduce very well the behavior of the optical weight obtained experimentally for the optimally doped BSCO in the work[2] and the *Figs.* 4*b* reproduce well the results for the overdoped BSCO[3]. We do not like to claim that result is the single correct explanation of the date obtained in[2,3]. We would like only to emphasize that the mere explanation of the temperature dependence of $W(T)$ observed in[2,3] does not necessitate to involve any unusual or exotic mechanisms of superconductivity. The results presented on *Figs.* 4*a*, 4*b* have been obtained to take in to account only the properties of normal state and very simple phonon spectra. We have performed the numerical calculations of $W(T)$ using mentioned above the more realistic phonon spectrum and the selfconsistent solution for the self-energy $\Sigma(\omega)$. The calculations have been done both for a normal and superconducting states and the results is shown on *Fig.* 5. In the total accordance with the discussion given in this work above the curves of $W(T)$ for the normal and superconducting states coincide with each other and cannot be discriminated on this *Fig.* The used phonon spectrum contains both phonon peaks at $\Omega_L \sim 200K$ and $\Omega_H \sim 400K$ and low temperature behavior of $W(T)$ can be easily changed by a redistribution of constant of the electron-phonon coupling between different peaks. The result for $W(T)$ obtained in the elastic scattering limit (*Eqs.* (47, 48, 52)) is also shown on the *Fig.* 5 and it coincides very well with that of given by the selfconsistent solution. We would like to emphasis that the real difference of the absolute values of these two curves smaller than 1%!

In the conclusion we shall discuss the results of numerical calculations of the temperature dependence of the restricted sum rule $W(\Omega_c, T)$ for various cutoff energies $\Omega_c$ which are presented on *Fig.* 6. The functions $W(\Omega_c, T)$ were calculated for a normal state only. For these calculations we have used the expression for the conductivity obtained by generalization of well known Nam's formula[27] on a case of a finite band. The explicit form of this expression for normal and superconducting states will be presented in our following publication. We present on *Fig.* 6 the result of the numerical calculations of the dependences of $W(\Omega_c, T)$ on $T^2$ for cutoff energies $\Omega_c$: 5; 100; 400; 1000; 2000; 5000 and 10000 $sm^{-1}$ (solid lines, from bottom to top, correspondingly) and the calculations in the purely elastic scattering limit (see *Eq.* (57)) with $\Gamma(T)$ from *Eq.* (52) (dotted line). (The corresponding curves for $\Omega_c = 20000\ sm^{-1}$ are presented on *Fig.* 5). This *Fig.* shows that the $T^2$ behavior of $W(\Omega_c, T)$ exists only for $\Omega_c \gg \Omega_{ph}$ and the slope of this line decreases with the increasing of $\Omega_c$. It is in a good agreement with the result obtained in the work[8]. In the area of the applicability of the elastic scattering approximation $\Omega_c \gg \Omega_{ph}$ this approach reproduces the exact curves with the asymptotic accuracy $\sim \left(\frac{\Gamma}{\Omega_c}\right)^2 \ll 1$ (not larger then 5%). These approximating curves $w(\Omega_c, T)$ with the same accuracy can be presented by the simple formula $w(\Omega_c, T) \approx 1 - \frac{2\Gamma(T)}{\Omega_c}$ (see *Eq.*57). We can calculate also the spectral weight temperature coefficient $B(\Omega_c)$ (see *Eq.*3) using the expression

$$B(\Omega_c) \approx \frac{1}{\tilde{\omega}_{pl}^2} \frac{\partial W(\Omega_c, T^2)}{\partial (T^2)}\bigg|_{T=150K} \tag{60}$$

The behavior $B(\Omega_c)$ is presented on *Fig.* 7 where the results of our selfconsistent calculations are designated as open circles. Also the results of the asymptotical approximation for Einstein spectra with $\Omega_E = 200K$ and $\Omega_E = 400K$ (see *Eq.*(59)) is presented (solid lines) on this *Fig.* It can be seen that these curves are also very close to experimental data[8]. We would like to underline here that there is a large difference between the values of $B(\Omega_c)$ at small $\Omega_c$ for the cases $\Omega_E = 200K$ and $\Omega_E = 400K$ in a large analogy with the results obtained in the work[8] for the overdoped and underdoped LSCO.

All these results allow us to assume that the types of the temperature dependence of spectral weight observable experimentally are determined by electron-phonon relaxation rate and connected, in fact, with the position of the phonon's modes and constants coupling of electrons with these modes. Therefore these data cannot contain the relevant information on the mechanism of superconductivity in HTCP.

**Acknowledgments.** This work was supported by RFBR under grants No 03-02-16252, No 04-02-17367, and grant NWO-RFBR No 047.016.005.

# References

1. R. Kubo, J. Phys. Soc. Japan **12**, 570 (1957)
2. D.N. Basov, et al., Science **283** (1999) 49.
3. H.J.A. Molegraaf, et al., Science **295** (2002) 2239.
4. A.F. Santander-Syro, et al., Europhys. Lett. **62**, 568 (2003)
5. A.F. Santander-Syro, Phys. Rev. **B 70,** 134504 (2004)
6. C.C. Homes et al., Phys. Rev. **B 69**, 024514 (2004).
7. A. V. Boris et al., Science **304,** 708 (2004).
8. M. Ortolani et al., Phys. Rev. Lett. **94,** 667002 (2005)
9. J.E. Hirsch, Science **295**, 2226 (2002).
10. D. van der Marel, in *Concepts in Electron Correlation*, Eds. A. Newson, V. Zlatio, Kluwer (2003), p. 7.
11. D. van der Marel, in *Strong Interaction Electrons in Low Dimensions*, Eds. D. Baeriswyl and L. De Giorgi, Kluwer Academic Publishers (200
12. M.R. Norman and C. P´epin, Phys. Rev. **B 66**, 100506 (2002)
13. A.E. Karakozov et al., Solid State Commun. **124** (2002) 708
14. J.P. Carbotte, E. Schachinger, Phys. Rev. **B 69**, 224501 (2004)
15. L. Benfatto et al., Phys. Rev **B 71**, 104511 (2005)
16. R.A. Ferrel, R.E. Glover Phys. Rev. **109,** 1398 (1958)
17. M. Tinkham, R.A. Ferrel, Phys. Rev. Lett. **2,** 331 (1959)
18. M. Tinkham, *Introduction to Superconductivity*, McGraw-Hill, New York, 1996.
19. G.M. Eliashberg, Sov. Phys. JETP, **11**, 696 (1960)
20. P. Monthoux, A.V. Balatsky, D. Pines, Phys. Rev. Lett. **67** (1991) 3448
21. T. Holstein, Ann. Phys. (N.Y.) **29**, 410 (1964)
22. J. R. Schrieffer, *Theory of superconductivity,* Addison Wesley (1988)
23. E.G. Maksimov, UFN **170** (2000) 1033; Physics-Uspekhi **43** (2000) 965
24. P.B. Allen, B. Mitrovic, in *Solid State Physics* Eds. H. Ehrenreich, F. Zeitz, D. Turnbull (Acad. Press, New York) v. **37** (1982)
25. J. Hubbard, Proc. Roy. Soc. (London) A281, 401 (1964)
26. H. Ehrenreich, L. Schwartz, in *Solid State Physics* Eds. H. Ehrenreich, F. Zeitz, D. Turnbull (Acad. Press, New York) v. **31** (1976)
27. S.B. Nam, Phys. Rev. **156** (1967) 470

# 1. FIGURE CAPTION

Fig.1. Normalized functions $G(z) = g(z)/N_b(\mu)$ in one-band interacting system, bare band $G_b(z) = g_b(z)/N_b(\mu)$ and approximation (48), (52) for $z = \omega/W_b$ at $T = 200K$.

Fig.2. Comparison of function $\omega[Z(\omega)-1]$ in one-band interacting system (solid lines) with corresponding functions for homogeneous electron gas (dotted lines) and approximation (49), (52) (open squares) at $T = 200K$.

Fig.3. Mass renormalization function $\operatorname{Re} Z(\omega) - 1$ in one-band interacting system at $T = 10K$.

Fig.4. The simplest approximation for optical spectral weight $w_{\mathrm{mod}}$ (*Eq.* 59) for Einstein phonon spectra with $\lambda = 1.5$ and $\Omega_E = 200K$ (*Fig.* 4a); $\Omega_E = 400K$ (*Fig.* 4b).

Fig.5. Normalized optical spectral weight in one-band interacting system (triangles) and approximation function (56), (52) (open circles).

Fig.6. Temperature dependences $W(\Omega_c, T^2)$ in one-band interacting system (solid lines) and its elastic scattering approximations $w(\Omega_c, T^2)$ (dotted lines) for cutoff energies $\Omega_c$: 5; 100; 400; 1000; 2000; 5000 and 20000 $sm^{-1}$ (from bottom to top, correspondingly).

Fig.7. Cutoff energy $\Omega_c$ dependence $B(\Omega_c)$ in one-band interacting system (dashed line with open circles) and $B(\Omega_c)$ for Einstein phonon spectra with $\lambda = 1.5$, $\Omega_E = 200K$, and $\Omega_E = 400K$ (solid lines).

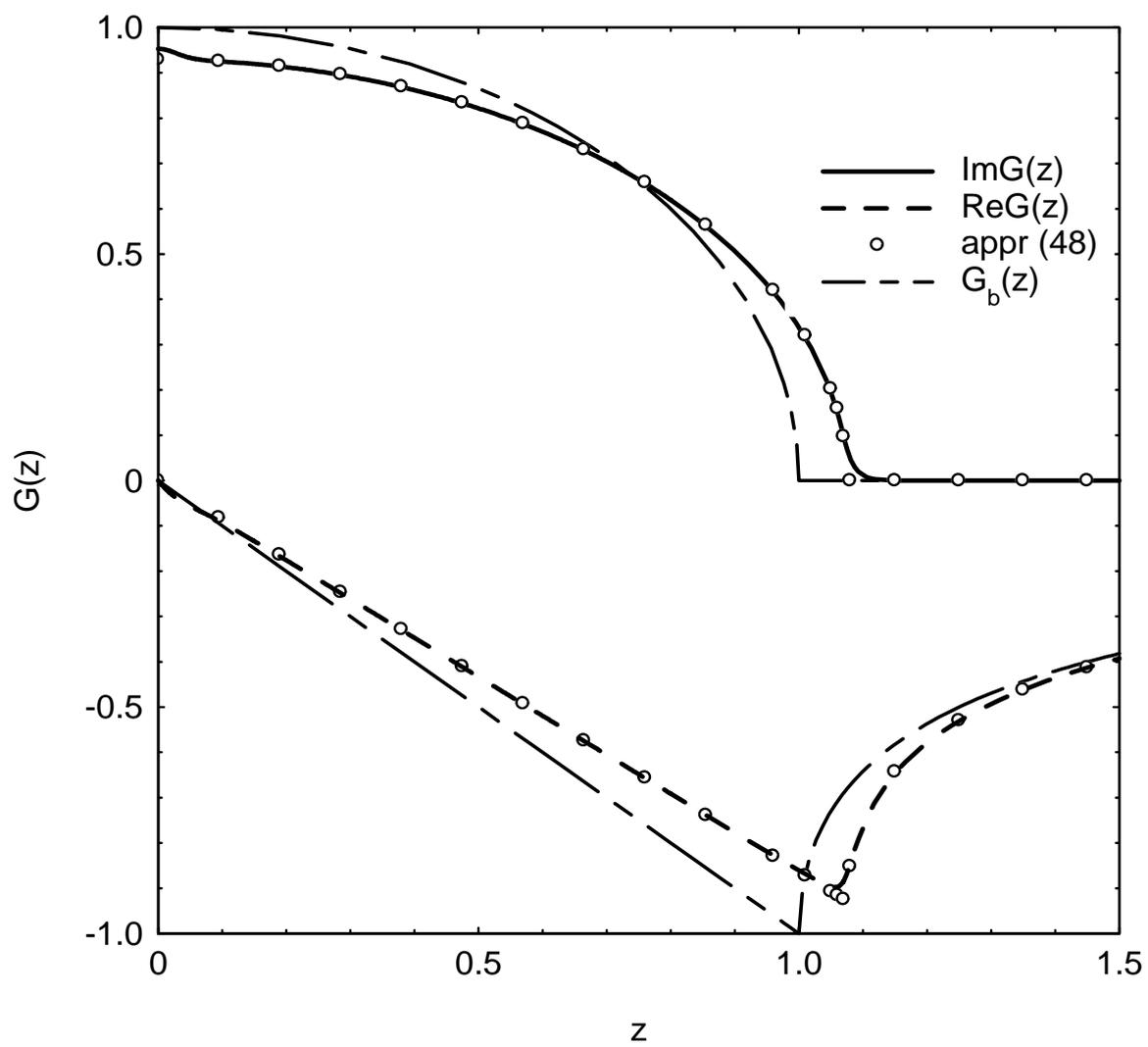

*Fig. 1.*

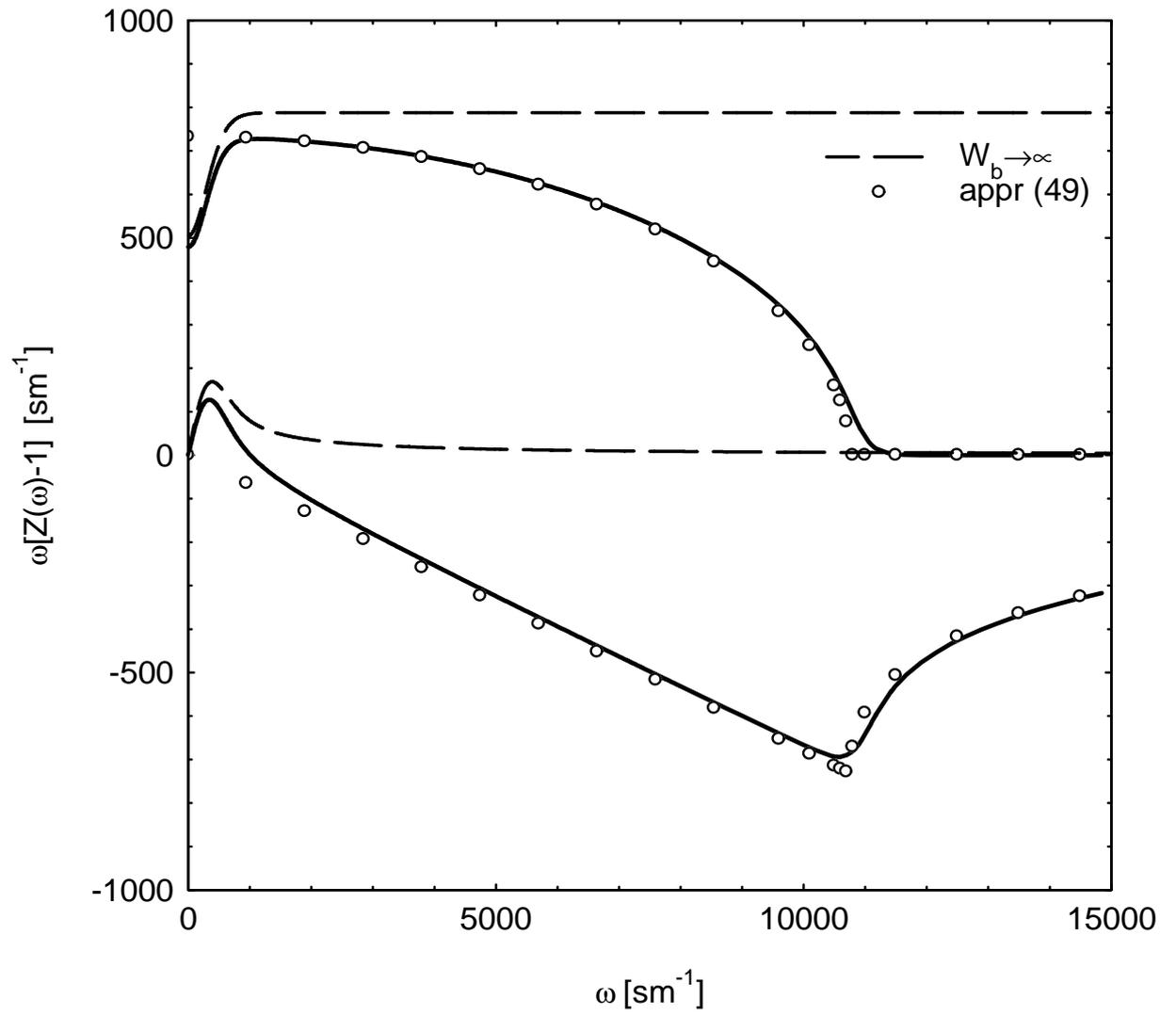

*Fig. 2.*

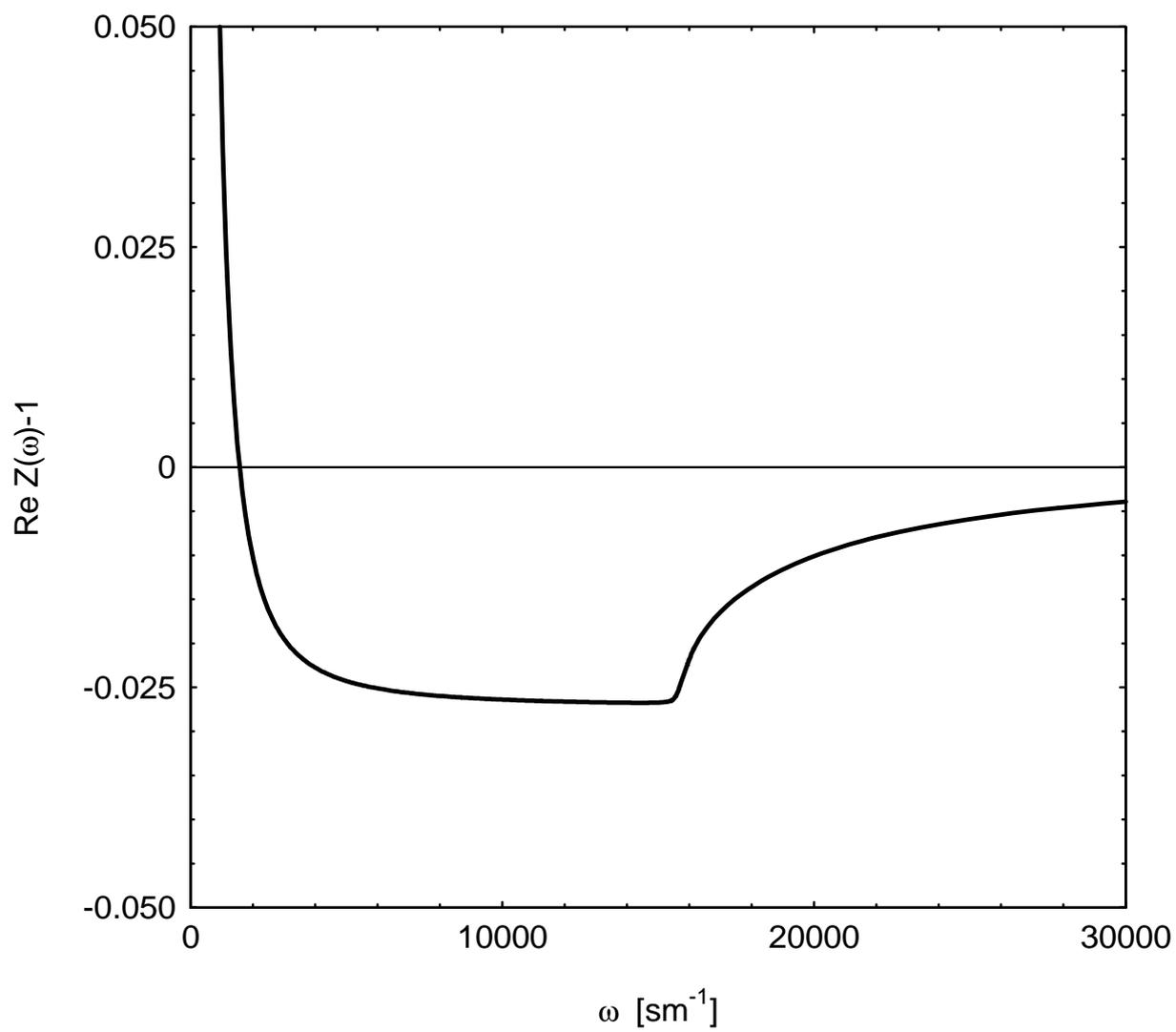

*Fig. 3.*

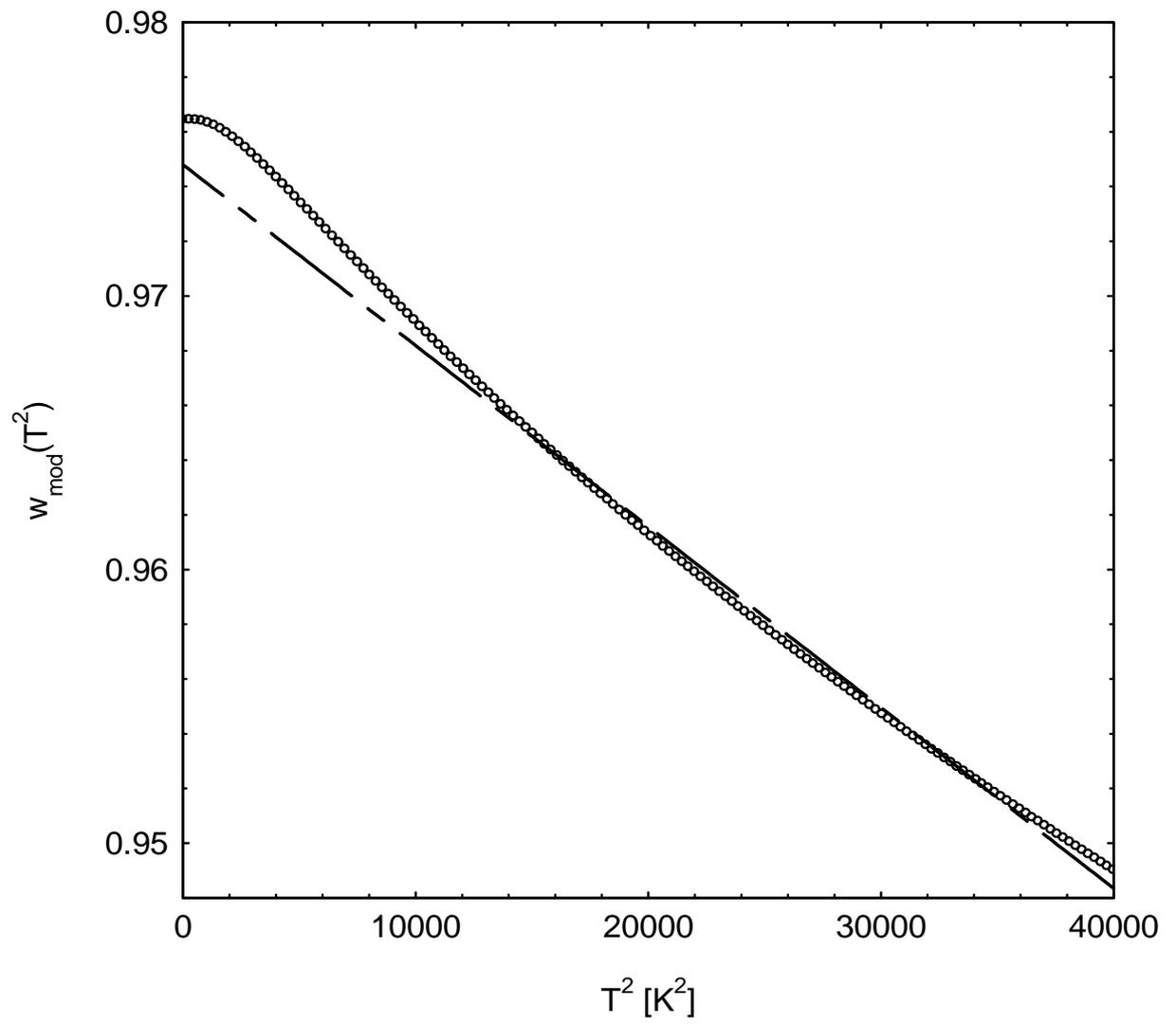

*Fig. 4a.*

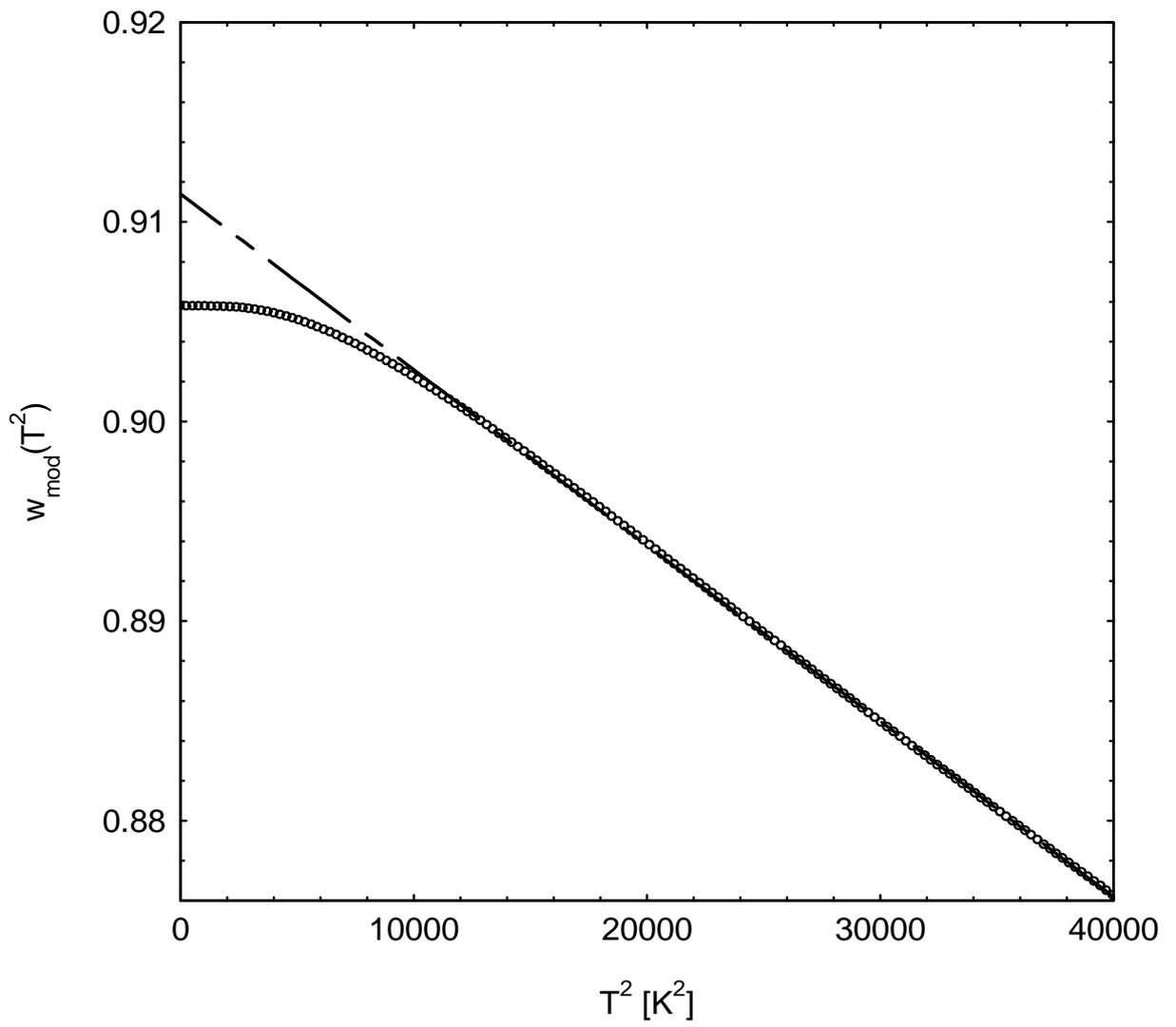

*Fig. 4b.*

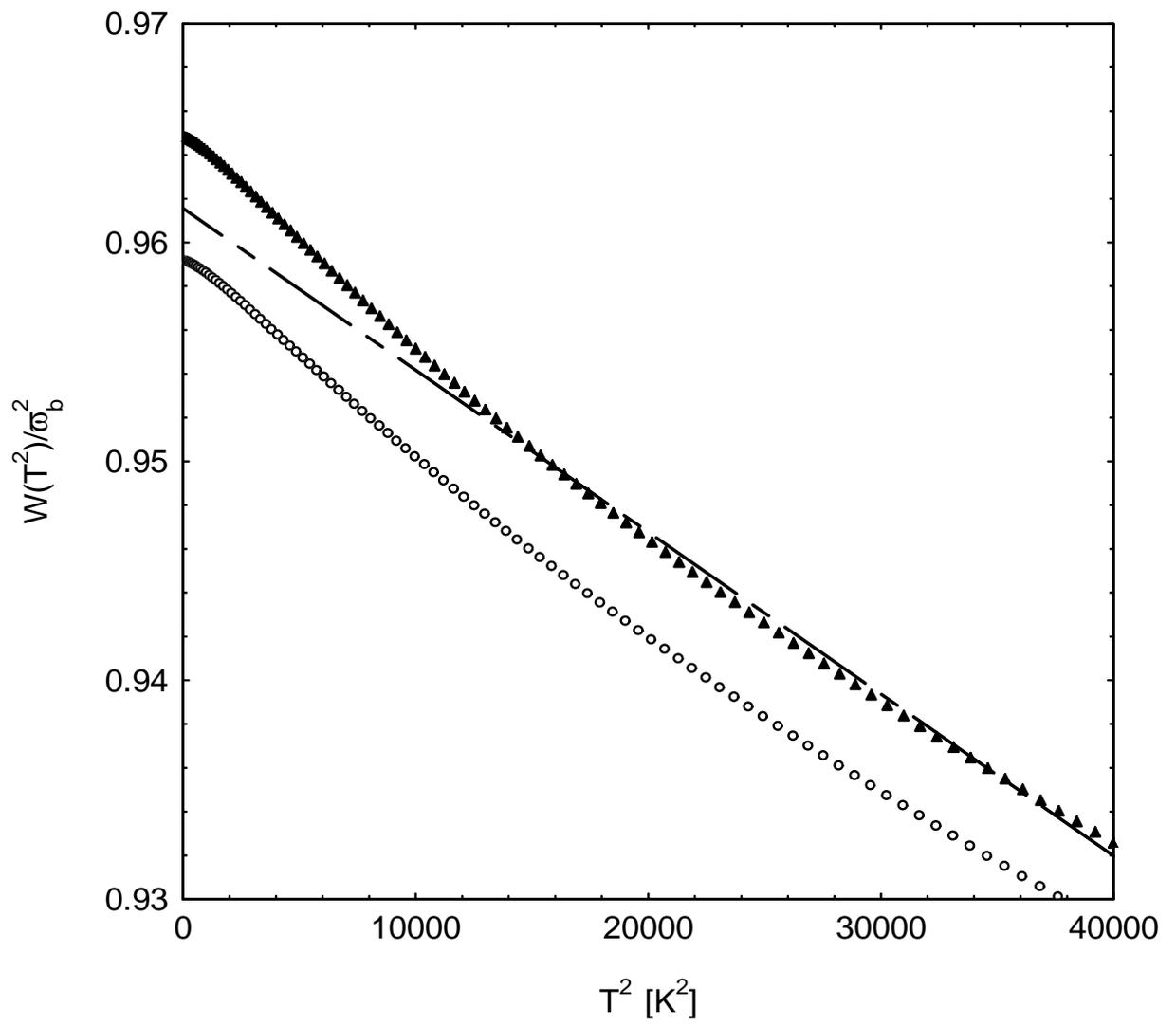

*Fig. 5.*

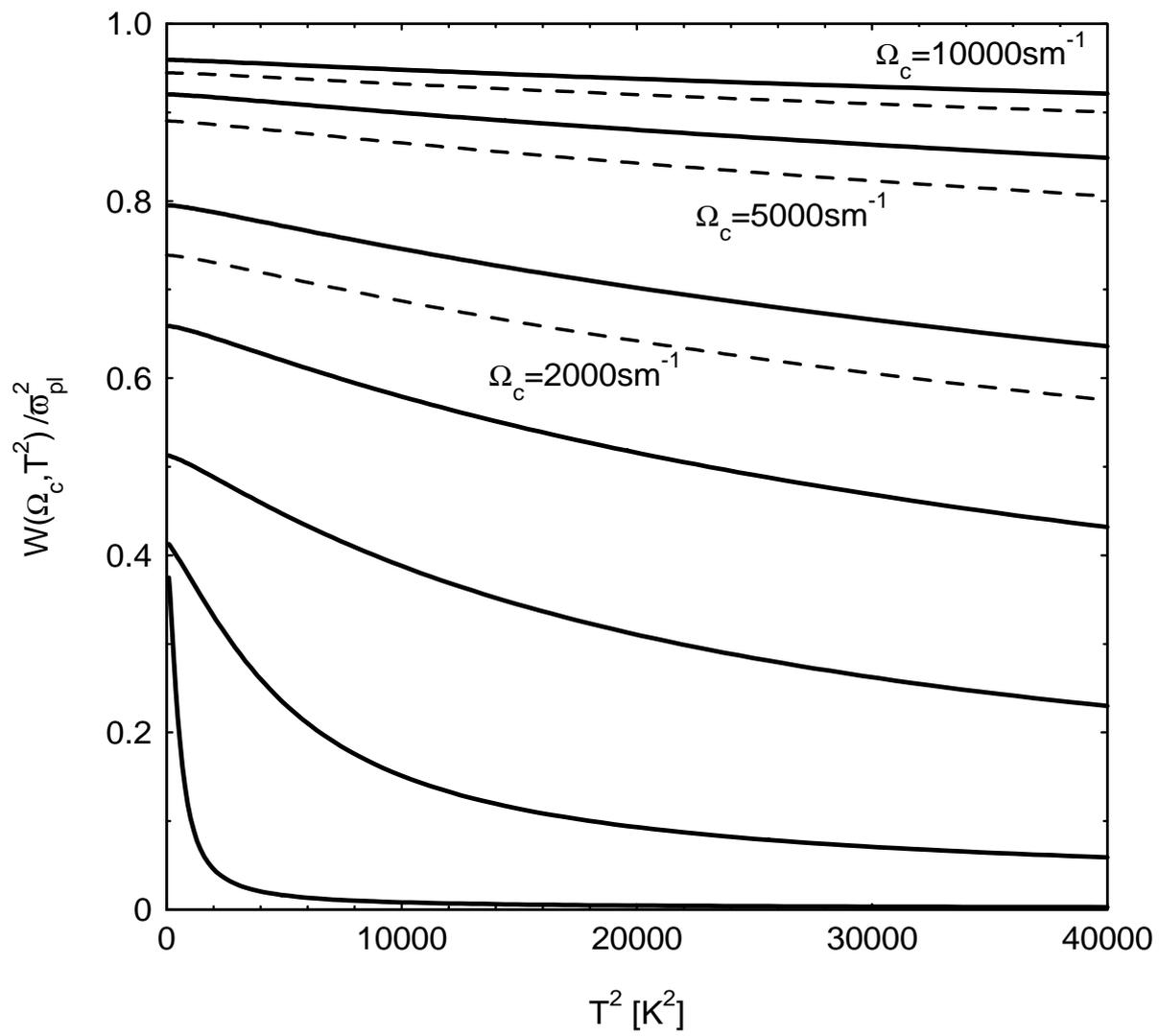

*Fig. 6.*

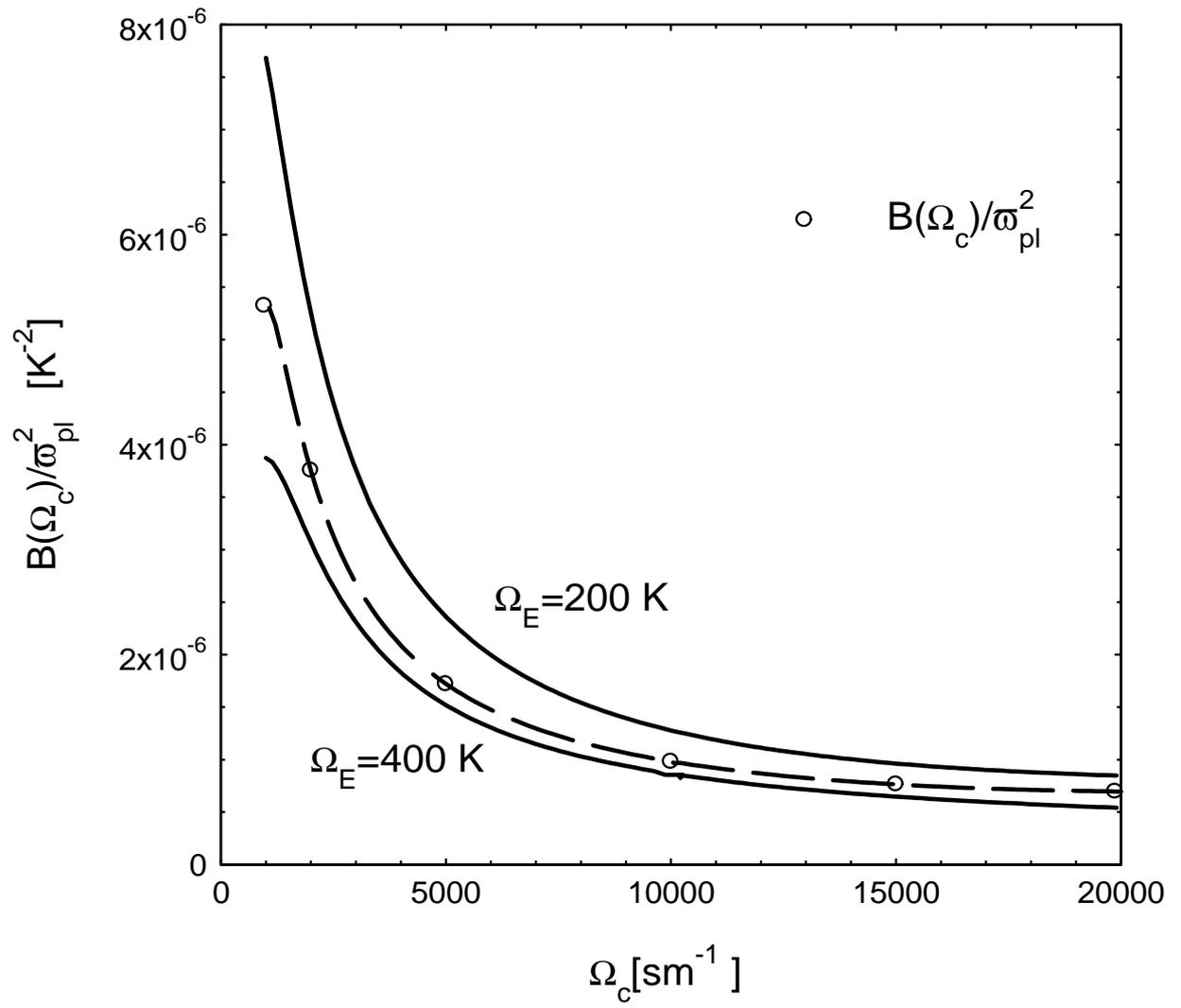

*Fig. 7.*